\documentclass[english,prl,preprint,amsfonts,amssymb,showpacs]{revtex4}
\usepackage[T1]{fontenc}
\usepackage[latin1]{inputenc}
\usepackage{amsmath}
\usepackage{babel}
\usepackage{graphics}

\makeatletter

\providecommand{\LyX}{L\kern-.1667em\lower.25em\hbox{Y}\kern-.125emX\@}

\makeatother

\begin{document}

\pacs{11.25.Yb, 98.80.Jk, 04.50.+h}

\preprint{hep-th/0304089}

\preprint{DAMTP-2003-29}

\title{Accelerating Cosmologies and a Phase Transition in M-Theory}

\author{Mattias N.R. Wohlfarth}

\email{M.N.R.Wohlfarth@damtp.cam.ac.uk}

\affiliation{Department of Applied Mathematics and Theoretical Physics, Centre for Mathematical
Sciences, University of Cambridge, Wilberforce Road, Cambridge CB3 0WA, U.K.}

\begin{abstract}
M-theory compactifies on a seven-dimensional time-dependent hyperbolic or flat
space to a four-dimensional FLRW cosmology undergoing a period of accelerated
expansion in Einstein conformal frame. The strong energy condition is violated
by the scalar fields produced in the compactification, as is necessary to evade
the no-go theorem for time-independent compactifications. The four-form field
strength of eleven-dimensional supergravity smoothly switches on during the
period of accelerated expansion in hyperbolic compactifications, whereas in
flat compactifications, the three-form potential smoothly changes its sign.
For small acceleration times, this behaviour is like a phase transition of the
three-form potential, during which the cosmological scale factor approximately
doubles. 
\end{abstract}

\date{\today}

\maketitle
At present, the study of accelerating cosmologies deserves particular interest
for several reasons. Recent astronomical measurements on certain types of supernovae,
as reported, e.g., in \cite{Rie01}, seem to indicate that our universe is not
only expanding, which has been known for a long time, but is undergoing an accelerated
expansion. Moreover, the hypothesis of another, inflationary, epoch of accelerated
expansion in the distant past is supported by measurements on the cosmic microwave
background, see, e.g., \cite{Ben03}. Phenomenologically motivated cosmological
models which exhibit accelerated expansion for at least a certain period of
time are easily constructed. The task, however, of deriving such a model from
a compactification of a higher-dimensional theory such as superstring or M-theory,
which are believed to incorporate the standard model consistently, is a more
difficult one. This is due to the existence of a no-go theorem for compactifications
on \textit{time-independent} internal spaces \cite{Gib85,MaNu01}. This theorem
states that if the strong energy condition is satisfied for a higher-dimensional
theory, which it is for ten or eleven-dimensional string or M-theory, then it
is also valid for the compactified theory. But then the time-time component
of the four-dimensional Ricci tensor is subject to the condition \( R_{00}\ge 0 \),
and, since \( R_{00} \) is related to the acceleration of the scale factor
\( S \) in an FLRW universe by \( R_{00}=-3\ddot{S}/S \), it follows that
acceleration is forbidden. 

A recent letter demonstrated the possibility of evading
this theorem \cite{ToWo03}; a solution of the vacuum Einstein
equations was presented for which spacetime is compactified on a
compact hyperbolic manifold
of \textit{time-varying} volume to a four-dimensional homogeneous and isotropic
FLRW universe with a phase of accelerated expansion. This work is easily extended
to the M-theory case where, in the low energy limit, the equations of motion
follow from the bosonic part of the action of eleven-dimensional supergravity
\cite{CJS78},
\begin{equation}
\label{eq. elaction}
S=\int _{11}\sqrt{-g}\left( R-\frac{1}{2\cdot 4!}F_{4}^{2}\right) ,
\end{equation}
involving the curvature scalar \( R \) and the four-form field strength \( F_{4}=dC_{3} \).
In principle, this action should additionally contain the Chern-Simons term
\( C_{3}\wedge F_{4}\wedge F_{4} \). This term, however, becomes irrelevant
below when \( F_{4} \) is realized as a volume form. The fact that \( F_{4} \)
can be non-zero in toroidal or spherical compactifications has been established
before \cite{ANT80,FrRu80}. One new point here is to look at the remaining possibility
and to investigate the effects of a non-zero \( F_{4} \) in a hyperbolic compactification. 

Consider the following eleven-metric, parameterized by functions of time \( K(t),\, L(t) \)
and \( S(t) \), 
\begin{equation}
\label{eq. elmetric}
ds_{11}^{2}=\left( \frac{K(t)}{L(t)}\right) ^{-7/6}ds_{E}^{2}+\left( \frac{K(t)}{L(t)}\right) ^{1/3}ds_{7}^{2}\, ,
\end{equation}
where
\begin{equation}
\label{eq. Emetric}
ds_{E}^{2}=-S(t)^{6}dt^{2}+S(t)^{2}d{\bf x}\cdot d{\bf x}
\end{equation}
is a four-dimensional FLRW metric in Einstein conformal frame. The internal
space with the metric
\begin{equation}
ds_{7}^{2}=g_{7mn}dy^{m}dy^{n}
\end{equation}
is taken to be a compact Einstein manifold with negative or zero curvature, such that
its Ricci tensor satisfies
\begin{equation}
R_{7mn}=-6\kappa ^{2}g_{7mn}\quad \textrm{for}\quad \kappa \ge 0\, .
\end{equation}
 The eleven-metric (\ref{eq. elmetric}), together with the four-form
\begin{equation}
\label{eq. F4}
F_{4}=bL(t)^{2}vol_{1,3}
\end{equation}
where \( vol_{1,3}=dt\wedge dx_{1}\wedge dx_{2}\wedge dx_{3} \), solves the
Einstein equations and the equation of motion of \( F_{4} \) derived from the
action (\ref{eq. elaction}) if
\begin{equation}
S(t)=K(t)^{7/12}L(t)^{-1/4}
\end{equation}
 and\begin{subequations}\label{eq. KL}
\begin{eqnarray}
K(t) & = & \left\{ \begin{array}{ccc}
\frac{\sqrt{21}\gamma }{14\kappa \sinh \frac{3}{7}\sqrt{21}\gamma |t|} & \quad \textrm{for}\quad  & \kappa >0\\
e^{\frac{3}{7}\sqrt{21}\gamma t} & \quad \textrm{for}\quad  & \kappa =0
\end{array}\right. ,\\
L(t) & = & \frac{3\gamma }{b\cosh 3\gamma t}\, .
\end{eqnarray}
\end{subequations}Note that not all constants of integration are displayed:
different time shifts may be chosen in \( K(t) \) and \(
L(t)\). Solutions that are of similar type have been
considered before, however, without appreciating their possible
accelerating behaviour, see e.g. \cite{Dem85}. Moreover, the specific
solution presented above
has a close relation to \( S \)-branes, see \cite{Oht03, Roy03} (and references
therein). It describes, in fact, an electrically
charged \( SM2 \)-brane, with a hyperbolic or flat transverse space, whose
three spatial worldvolume directions correspond to the spatial sections of the
four-dimensional FLRW cosmology considered above. It is clear from \( K(t) \)
in (\ref{eq. KL}) that the \( \kappa >0 \) solution with hyperbolic internal
space does not simply reduce to the one for \( \kappa =0 \) where the internal
space is flat. The same happens for the field strength: the solution for \( L(t) \)
with non-zero field strength parameter \( b\neq 0 \) does not simply reduce
to the one for \( b=0 \), given in \cite{ToWo03}, where instead \( L(t)=e^{3\gamma t} \)
(with \( \gamma  \) set to \( \gamma =1 \)).

\begin{figure}
{\par\centering \resizebox*{0.6\columnwidth}{!}{\includegraphics{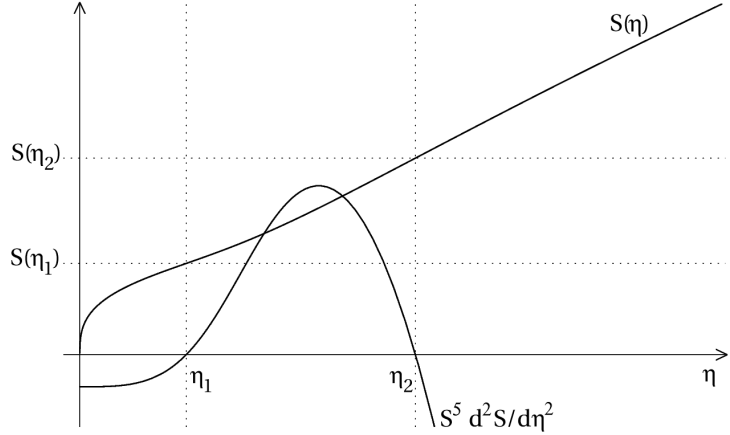}} \par}

\caption{\label{fig. figure1}\textit{Plot of a typical scale factor \protect\( S(\eta )\protect \)
and its acceleration \protect\( d^{2}S/d\eta ^{2}\protect \)
(multiplied by $S^5$). }}
\end{figure}

The standard time coordinate \( \eta  \) in Einstein frame is defined by \( d\eta =S(t)^{3}dt \),
see (\ref{eq. Emetric}). Expansion of the Einstein frame scale factor corresponds
to \( dS/d\eta >0 \) and acceleration to \( d^{2}S/d\eta ^{2}>0 \). Hence,
as in the zero field strength case \cite{ToWo03}, it is easily shown that the
above solutions admit a period of accelerated expansion for negative \( t \)
in a certain interval
\begin{equation}
\label{eq. accint}
\frac{t_{1}}{\gamma }<t<\frac{t_{2}}{\gamma } 
\end{equation}
(which maps into an interval $\eta_1<\eta<\eta_2$ of positive $\eta$), see figure \ref{fig. figure1}.
 Such accelerated expansion, however, can only be possible if the strong energy
condition is violated by the scalar fields arising in the compactification process.
These are the scale \( e^{M} \) of the internal space, which is determined
by
\begin{equation}
\label{eq. M}
M(t)=\frac{1}{6}\ln \frac{K(t)}{L(t)}\, ,
\end{equation}
and a term coming from \( F_{4} \). Reducing the eleven-dimensional action
(\ref{eq. elaction}) by substituting the metric ansatz (\ref{eq. elmetric}),
one finds a four-dimensional scalar Lagrangian
\begin{equation}
\mathcal{L}=7\nabla ^{2}_{E}M-\frac{63}{2}(\nabla _{E}M)^{2}-42\kappa ^{2}e^{-9M}-\frac{1}{2\cdot 4!}e^{21M}F_{4}^{2}\, .
\end{equation}
Note that this expression is given in Einstein frame and is only reached by
a suitable conformal transformation of the four-dimensional part of the original
eleven-metric. It is necessary to go to the Einstein frame in order to avoid
interpretational ambiguities: in particular, this frame is characterized by
a time-independent Newton constant. From the energy-momentum tensor \( T_{\mu \nu } \)
of \( \mathcal{L} \) one finds
\begin{eqnarray}
X & \equiv  & T_{\eta \eta }-\frac{1}{2}Tg_{E\eta \eta }\nonumber \\
 & = & 63{M'}^{2}-42{\kappa }^{2}e^{-9M}-\frac{1}{2}b^{2}e^{-21M}
\end{eqnarray}
where \( M'=dM/d\eta  \). This quantity is required to be positive by the strong
energy condition, and indeed,
\begin{equation}
X=-6S^{-1}\frac{d^{2}S}{d\eta ^{2}}\, ,
\end{equation}
such that the strong energy condition is violated if, and only if, the acceleration
of the scale factor \( S \) is positive. Note, from the second term in \( X \),
that a hyperbolic internal space, where \( \kappa >0 \), gives rise to a negative
contribution. For a flat internal space, this term vanishes and the strong energy
condition can be violated only by the third term which is due to the four-form
\( F_{4} \). The latter contributes negatively because of a more general observation
\cite{Gib85}: \( p \)-forms in \( n \) dimensions may violate the strong
energy condition for \( p\ge n-1 \). Such an effect, that the four-form of
eleven-dimensional supergravity gives rise to a negative cosmological constant
upon reduction to four dimensions, has also been noted in \cite{ANT80}.

\begin{figure}
{\par\centering \resizebox*{0.6\columnwidth}{!}{\includegraphics{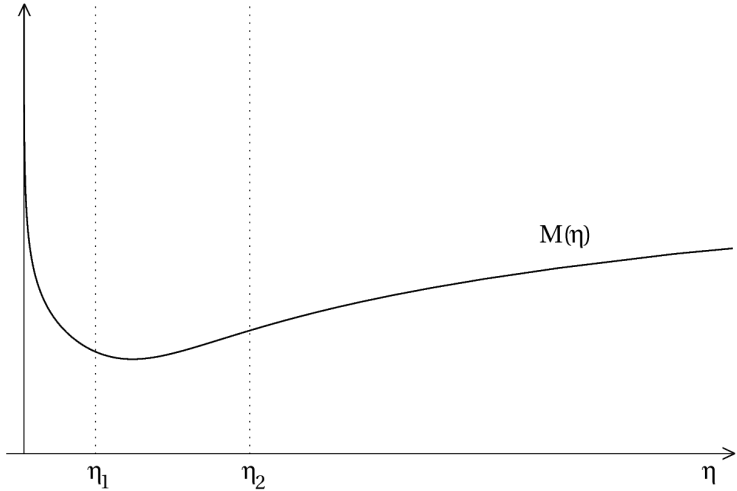}} \par}

\caption{\label{fig. figure2}\textit{The scale of the internal space given by \protect\( M(\eta )\protect \). }}
\end{figure}

The phase of acceleration has several interesting properties. In the following
they are discussed for a hyperbolic internal space. For a flat internal space
one proceeds similarly, and a summary of the results for this case is given
below. Now let \( \kappa >0 \). Accelerated expansion starts at the time \( t_{1}/\gamma  \)
and ends at \( t_{2}/\gamma  \) where one finds \( t_{1}\cong -0.78 \) and
\( t_{2}\cong -0.15 \) as the two negative solutions of the equation
\begin{eqnarray}
 & -(5-\sqrt{21})\cosh 6(1+\frac{1}{7}\sqrt{21})t+8\cosh \frac{6}{7}\sqrt{21}t & \nonumber \\
 & -(5+\sqrt{21})\cosh 6(1-\frac{1}{7}\sqrt{21})t+8\cosh 6t-10 & =0
\end{eqnarray}
which follows from \( d^{2}S/d\eta ^{2}=0 \). In terms of the physical time
variable \( \eta  \), the accelerating phase is bounded by
\begin{equation}
\label{eq. times}
\eta _{i}=A(t_{i})b^{3/4}\kappa ^{-7/4}\quad \textrm{for}\quad i=1,2
\end{equation}
where the function \( A(t) \) is given by
\begin{equation}
\label{eq. int}
A(t)=\left( \frac{3}{2^{14}7^{7}}\right) ^{1/8}\int _{-\infty }^{t}ds\frac{\cosh ^{3/4}3s}{\sinh ^{7/4}\frac{3}{7}\sqrt{21}|s|}\, .
\end{equation}
The difference \( \eta _{\textrm{acc}}=\eta _{2}-\eta _{1} \) gives the total
acceleration time
\begin{equation}
\label{eq. totime}
\eta _{\textrm{acc}}=[A(t_{2})-A(t_{1})]b^{3/4}\kappa ^{-7/4},
\end{equation}
and the expansion factor during the period of accelerated expansion is obtained
to be
\begin{equation}
a=\frac{S(\eta _{2})}{S(\eta _{1})}\cong 2.155\, .
\end{equation}
Note that the factor \( a \) is constant and independent of the parameters
of the solution (\ref{eq. KL}). This implies that there can be no hope of using
this solution in any inflationary scenario, where the required expansion to
solve cosmological issues, such as the horizon or the flatness problem, is of
the order of magnitude \( \mathcal {O}(e^{60}) \). The total acceleration time
\( \eta _{\textrm{acc}} \), on the other hand, depends on the field strength
parameter \( b \) as well as on the curvature \( \kappa  \) of the internal
space. It tends to zero for \( b\rightarrow 0 \) or \( \kappa \rightarrow \infty  \)
in which limit the scale factor doubles instantaneously. The scale of the internal
space is determined by the behaviour of \( M \) in (\ref{eq. M}). One finds
that \( M\rightarrow \infty  \) as \( \eta \rightarrow 0 \) and also as \( \eta \rightarrow \infty  \).
In the accelerating phase, \( M \) assumes a minimum value, as is shown in
figure \ref{fig. figure2}.
 Looking at the solution in the whole range of \( \eta  \) shows that there
is no stabilisation of the internal space scale, which seems to be a more general
problem in hyperbolic compactifications \cite{NSST02}. The fact, however, that
the scale has a minimum, which is reached very fast, means that there is a natural
'compactification' of the internal space until the end of the accelerating phase.
The following decompactification is a very slow process.

\begin{figure}
{\par\centering \resizebox*{0.6\columnwidth}{!}{\includegraphics{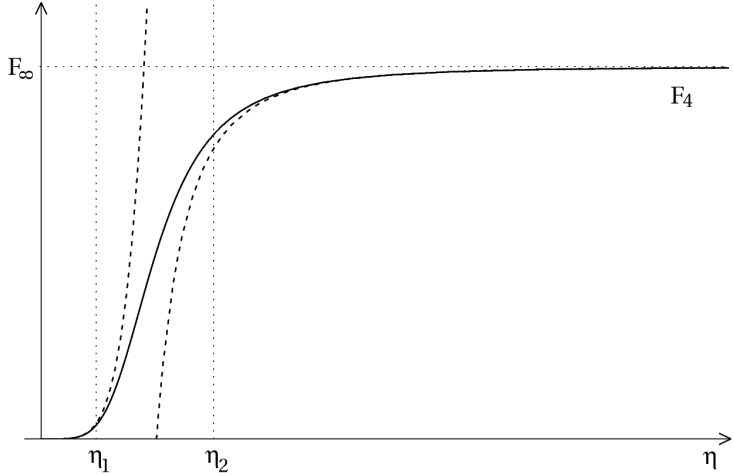}} \par}

\caption{\textit{\label{fig. figure3}The behaviour of the four-form \protect\( F_{4}\protect \)
as a function of \protect\( \eta \protect \) (for a hyperbolic internal space
\protect\( \kappa >0\protect \)). }}
\end{figure}

The four-form field \( F_{4} \) smoothly switches on in the phase of accelerated
expansion. For small times \( \eta \rightarrow 0 \), it behaves like
\begin{equation}
\label{eq. F4to0}
F_{4}\, \stackrel{\eta \rightarrow 0}{\sim }\, b^{-(1+2\epsilon )}\gamma ^{2}\kappa ^{14\epsilon /3}\eta ^{8\epsilon /3}vol_{1,3}
\end{equation}
where \( \epsilon =(\sqrt{21}/3-1)^{-1} \), whereas in the limit \( \eta \rightarrow \infty  \),
the field \( F_{4} \) tends to a constant value \( F_{4}=F_{\infty }vol_{1,3} \)
with
\begin{equation}
F_{\infty }=9\gamma ^{2}/b\, .
\end{equation}
More precisely,
\begin{equation}
F_{4}\, \stackrel{\eta \rightarrow \infty }{\sim }\, \left( F_{\infty }-9\gamma ^{2}b(6\kappa )^{-14/3}[\frac{3}{4}(\eta -\eta _{0})]^{-8/3}\right) vol_{1,3}\, ,
\end{equation}
see also figure \ref{fig. figure3}.
 As \( b\rightarrow 0 \), the four-form value at infinity \( F_{\infty } \)
may be kept constant by choosing an appropriate value for \( \gamma  \). But
this limit coincides with the one of zero total acceleration time. For small
\( b \), therefore, the three-form potential \( C_{3} \) of eleven-dimensional
supergravity is behaving as in a first order phase transition, characterized
by the fact that its derivative \( F_{4}=dC_{3} \) becomes discontinuous. The
four-form field \( F_{4} \) is almost instantaneously switched on to its \( \eta \rightarrow \infty  \)
value \( F_{\infty }vol_{1,3} \) while the scale factor \( S(\eta ) \) of
the FLRW universe jumps by the factor \( a\cong 2.155 \). 

\begin{figure}
{\par\centering \resizebox*{0.6\columnwidth}{!}{\includegraphics{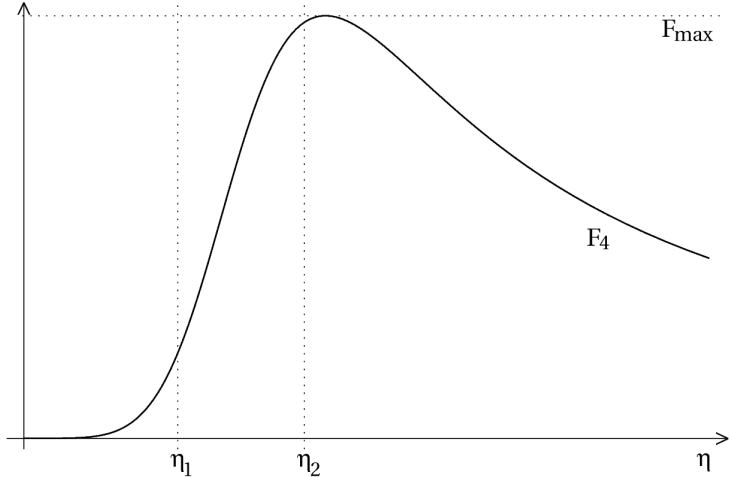}} \par}

\caption{\textit{\label{fig. figure4}The behaviour of the four-form \protect\( F_{4}\protect \)
as a function of \protect\( \eta \protect \) (for a flat internal space \protect\( \kappa =0\protect \)). }}
\end{figure}

Many of the results for a flat internal space, where \( \kappa =0 \), are similar
to those for the hyperbolic case \( \kappa >0 \), some are very different.
The appropriate expressions (\ref{eq. times}, \ref{eq. totime}) for the times
which characterize the period of accelerated expansion are essentially obtained
by replacing \( \kappa  \) by \( \gamma  \) (the integral in (\ref{eq. int})
differs, of course). The expansion factor of the FLRW cosmology in this period
becomes \( a\cong 1.355 \). For the internal scale one again finds an evolution
as shown in fig. \ref{fig. figure2}. The interesting changes appear in the
behaviour of \( F_{4} \). While (\ref{eq. F4to0}) stays the same up to the
replacement \( \kappa \mapsto \gamma  \), one finds that \( F_{4} \) tends
to zero also for \( \eta \rightarrow \infty  \). For some finite \( \eta _{\textrm{max}} \),
the four-form assumes a maximum value which equals the limiting value of the
\( \kappa >0 \) case, \( F_{\textrm{max}}=F_{\infty } \). For a plot of \( F_{4} \)
see figure \ref{fig. figure4}.
 These differences between \( \kappa >0 \) and \( \kappa =0 \) are best understood
from the three-form potential
\begin{equation}
C_{3}=\frac{3\gamma }{b}\tanh 3\gamma t\, dx_{1}\wedge dx_{2}\wedge dx_{3}\, .
\end{equation}
 In the hyperbolic case \( \kappa >0 \), a negative \( C_{3} \) at \( \eta =0 \)
(\( t=0 \)) is switched off as \( \eta \rightarrow \infty  \) (\( t\rightarrow 0 \)).
The value \( F_{\infty } \) of \( F_{4} \) is the slope at \( t\rightarrow 0 \).
The difference for \( \kappa =0 \) is that, here, \( \eta \rightarrow \infty  \)
corresponds to \( t\rightarrow \infty  \) because \( S(t) \) is non-singular.
Thus, a negative \( C_{3} \) at \( \eta =0 \) (\( t=0 \)) makes a transition
to a positive \( C_{3} \) of the same absolute value as \( \eta \rightarrow \infty  \)
(\( t\rightarrow \infty  \)). This transition takes place around \( \eta _{\textrm{max}} \).
It has the slope \( F_{\textrm{max}} \) and can be made discontinuous in the
limit \( b\rightarrow 0 \) while keeping the ratio \( \gamma /b \) constant. 

The alignment of the four-form along the directions \( (0\, 1\, 2\, 3) \) of
the final cosmology is not the only possibility consistent with the rotational
symmetry of FLRW spacetimes. Rather than choosing \( F_{4} \) as above, one
might also consider aligning it along \( (0\, |\, 4\, 5\, 6) \), \( (1\, 2\, 3\, |\, 4) \)
or along directions \( (4\, 5\, 6\, 7) \) completely in the internal space.
Another interesting question is whether there are solutions where the Chern-Simons
term of the eleven-dimensional supergravity action becomes important, or whether
it is possible to find solutions where the internal space is a product of several
manifolds. With such extensions, one might hope to be able to control the properties
of the phase of accelerated expansion or even to combine several of these phases
into a single solution. This then might provide a key to derive more realistic
cosmologies, including inflation, from M-theory.
After completion of this work, several papers have appeared which
explore further aspects of accelerating cosmologies from string and
M-theory compactifications \cite{EmGa03, Oht03a, CHNW03}.

\medskip

\begin{acknowledgments}

\textit{Acknowledgments.} MNRW thanks Paul K. Townsend for valuable discussions.
He gratefully acknowledges financial support from the Gates Cambridge Trust.

\end{acknowledgments}

\end{document}